\documentstyle[twocolumn,prb,aps,epsf,psfig]{revtex}

\begin{document}
\title{\bf Simulation of a cusped bubble rising in a viscoelastic
fluid with a new numerical method}
\author{A.J. Wagner, L. Giraud\thanks{Centre for Systems Engineering and Applied Mechanics (CESAME),
      Universite catholique de Louvain,
      Batiment Euler, Av. Georges Lemaitre 4,
      B-1348 Louvain-la-Neuve, Belgium},
 and C.E. Scott\\
}
\address{
Department of Materials Science and Engineering,
 Massachusetts Institute of Technology,
77 Massachusetts Avenue., Cambridge, MA 02139, U.S.A.
}
\date{\today}
\maketitle
\begin{abstract}
We developed a new lattice Boltzmann method that allows the simulation
of two-phase flow of viscoelastic liquid mixtures. We used this new
method to simulate a bubble rising in a viscoelastic fluid and were
able to reproduce the experimentally observed cusp at the trailing end of
the bubble.\\
\\
\end{abstract}
%


\section{Introduction}
The study of viscoelastic fluids is of great scientific interest and
industrial relevance. Viscoelastic fluids are fluids that show not
only a viscous flow response to an imposed stress, as do Newtonian
fluids, but also an elastic response. Viscoelastic effects are almost
universally observed in polymeric liquids\cite{bird}, where they often
dominate the flow behavior. They can also be observed in simple
fluids, especially in high frequency testing\cite{boon} or in
under-cooled liquids\cite{goetze}. Because most research into
viscoelastic liquids, especially that with an eye toward engineering
applications, is directed toward polymeric liquids, the viscoelastic
behavior of simple liquids is not as well known among researchers. The
fact that the manifestation of viscoelasticity does not require the
presence of polymer molecules is at the heart of our approach, as will
become clear in the description of the viscoelastic model.

Although in most practical problems involving polymeric materials the
viscosities of the materials involved are so large that the creeping
flow approximation is valid, the non-linearity introduced by the
viscoelastic response of the liquid makes it difficult to treat any
but the most simple cases analytically. In engineering applications
the situation is often further complicated by the fact that the system
is comprised of several immiscible or partially miscible components
with different viscoelastic properties. Examples of this include
polymer blending, where two immiscible polymers are melted and mixed
in an extruder, and the recovery of an oil-and-water mixture from
porous bed rock. Simulation of these systems is very important, but
due to the complexities only few numerical approaches exist to
date. Boundary element methods have been used to simulate such systems
with varying degrees of success, but the allowable complexity of the
interface morphology is very limited in such approaches. Lattice
Boltzmann simulations have been shown to be very successful for
Newtonian two-component systems with complex interfaces\cite{PRL}, but
for viscoelastic fluids the lattice Boltzmann models, derived by
Giraud {et al.}\cite{giraud_epl,giraud}, are limited to one-component
systems.

\begin{figure}
\begin{center}
\begin{minipage}{4cm}
\centerline{\psfig{figure=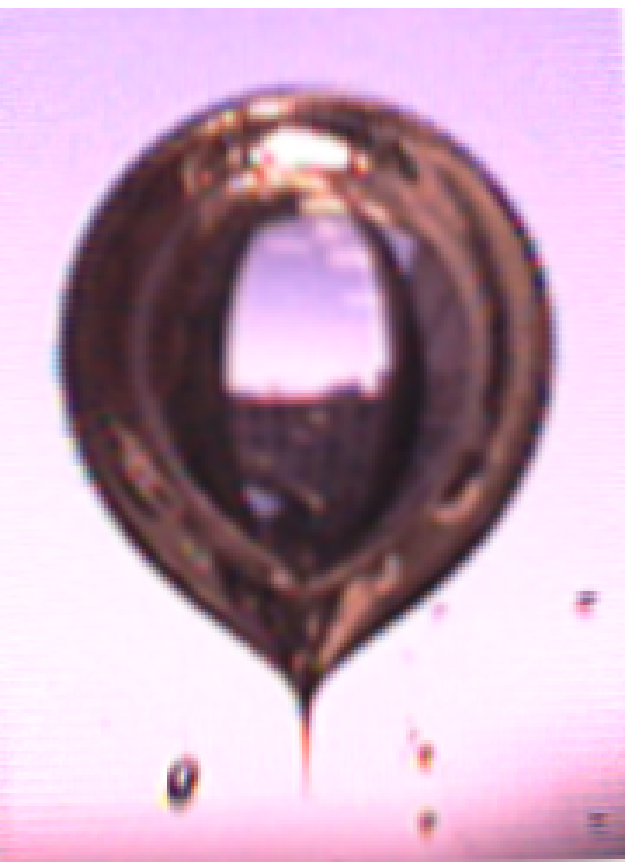,width=3.8cm}}
\begin{center} (a) \end{center}
\end{minipage}
\begin{minipage}{4cm}
\centerline{\psfig{figure=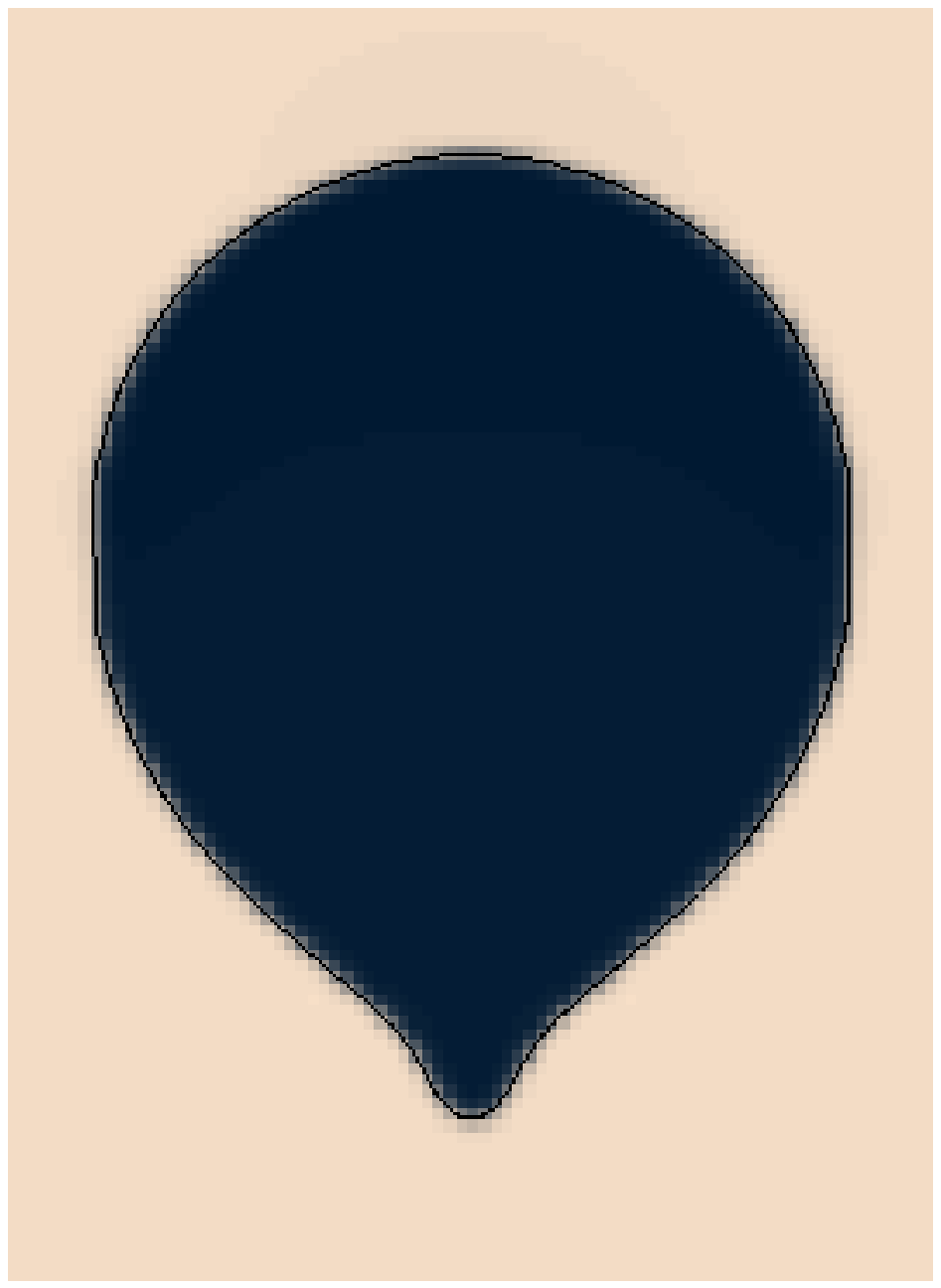,width=3.94cm}}
\begin{center} (b) \end{center}
\end{minipage}
\end{center}
\caption{ (a) An air bubble rising in $\mbox{Ivory}^{\bigcirc
\!\!\!\!\!\mbox{\tiny \rm R}}$ soap and (b) the simulation results for
a low viscosity drop rising in a Oldroyd B fluid on a 256x1024 lattice
for a drop of radius $R_0=35$. The simulated bubble has a cusp that is
rounded at the tip due to the finite thickness of the interface ($\sim 3$ lattice
spacings).}
\label{fig:Bubble}
\end{figure}

%

  In this article we report the successful combination of both
two-component and viscoelastic features into a two-dimensional
lattice Boltzmann model. We used this model to simulate a bubble rising
in a viscoelastic liquid (see Figure 1) and in this letter report the
first successful simulation of the experimentally observed cusp.

\section{Lattice Boltzmann}
We use a two-dimensional lattice Boltzmann model on a square lattice
with a velocity set of
 $\{{\bf v}_i\}=\{(0,0)$, $(0,1)$, $(1,0)$, $(0,-1)$, $(-1,0),$
 $(1,1),$ $(-1,1)$, $(-1,-1)$, $(1,-1)\}$ and a corresponding set of
 densities $\{ f_i \}$, but following Giraud {\it et al.}\cite{giraud}
we introduce two densities for each non-zero velocity. We use a BGK
lattice Boltzmann equation that contains the full collision matrix
$\Lambda_{ij}$ 
\begin{eqnarray} \label{eqn:lb}\label{feqn}
f_i(x+v_i \Delta t, t+\Delta t) &=& f_i(x,t)\nonumber\\
&&+\Delta t\Lambda_{ij} (f_j^0(x,t)-f_j(x,t))
\end{eqnarray}
where the summation rule for repeated indices is implied and the
required properties of the equilibrium distributions $f_i^0$ are
discussed below. The local density is given by $\rho=\sum_i f_i$ and
the momentum by $\rho {\bf u}=\sum_i f_i {\bf v}_i$.

In order to simulate a two-component mixture we define a second set of
nine densities, $\{ g_i\}$, with an appropriate equilibrium distribution,
$\{ g_i^0\} $. These densities represent the density difference of the
two components A and B as $\phi=\sum g_i=\rho_A-\rho_B$, where the
total density introduced earlier is $\rho = \rho_A +\rho_B$.
For the $g_i$s we choose a single relaxation time lattice Boltzmann equation
\begin{eqnarray}
  g_i(x+v_i \Delta t,t+\Delta t)&=&g_i(x,t)\nonumber\\
&&+\frac{\Delta t}{\tau}
  (g_i^0(x,t)-g_i(x,t)),\label{geqn}
\end{eqnarray}
where $\tau$ is the relaxation time and $g_i^0$ is the equilibrium
distribution. 

To use the lattice Boltzmann method in order to simulate fluid flow,
mass and momentum conservation have to be imposed. Mass and momentum
conservation are equivalent to constraints on the equilibrium
distributions:
\begin{equation}
\sum_i f_i^0=\rho,\;\;\;
\sum_i g_i^0= \phi,\;\;\;
\sum_i f_i^0 {\bf v}_i=
\rho {\bf u}. \label{momentum}
\end{equation}
There will be further constraints on the permissible equilibrium
distributions in order for the corresponding macroscopic equations to
be isotropic and to simulate the systems in which we are
interested. In the next two subsections we will summarize the physics
that we want to incorporate and then we will discuss how it imposes
constraints on the equilibrium distributions and eigenvalues.

\subsection{Binary mixtures}
To simulate a binary mixture we follow the approach of Orlandini {\it
et al.} \cite{enzo} and begin with a free energy functional $\Psi$
that consists of the free energy for two ideal gases and an
interaction term as well as a non-local interface term:
\begin{eqnarray}
\Psi[\rho_A,\rho_B]&=&\int_{\bf x} \left[T\rho_A \ln(\rho_A)
+ T\rho_B \ln(\rho_B)\nonumber\right.\\
&&\left. + \lambda \rho_A \rho_B+\kappa 
|\partial_{\bf x} (\rho_A-\rho_B)|^2
\right] d{\bf x},
\end{eqnarray}
where the densities $\rho_A$ and $\rho_B$ are functions of ${\bf
x}$. The repulsion of the two components is introduced in the
$\lambda$ term and $\kappa$ is a measure of the energetic penalty
for an interface.
When we write this free energy functional in terms of the total
density, $\rho$, and the density difference, $\phi$, we can derive the
chemical potential, $\mu$, and the pressure tensor, $P_{\alpha\beta}$,
as\cite{shear}:
\begin{eqnarray}
\mu&=&\frac{\delta \Psi}{\delta \phi}=\partial_\phi \Psi -\kappa
\partial_\gamma \partial_\gamma \phi,\\
P_{\alpha\beta} &=& (\rho \partial_\rho \Psi+ \phi \partial_\phi
\Psi)\delta_{\alpha\beta} \nonumber \\
&& + \kappa (\partial_\alpha \phi \partial_\beta \phi -\frac{1}{2}
\partial_\gamma \phi \partial_\gamma \phi \delta_{\alpha\beta} 
- \phi \partial_\gamma \partial_\gamma \phi \delta_{\alpha\beta}),
\end{eqnarray}
where $\delta$ indicates a functional derivative and
$\delta_{\alpha\beta}$ is the Kronecker delta. For a
two-component model we fix the further moments of the equilibrium
distributions\cite{shear}:
\begin{eqnarray}
\sum_i g_i^0 {\bf v}_i &=& \phi {\bf u},\\
\sum_i f_i^0 v_{i\alpha}v_{i\beta}&=&P_{\alpha\beta} + \rho u_\alpha u_\beta,\\
\sum_i g_i^0 v_{i\alpha}v_{i\beta}&=&\mu \delta_{\alpha\beta}
+\phi u_\alpha u_\beta. \label{pab}
\end{eqnarray}
Thus far, the model 
allows us to simulate a binary mixture that phase separates
below a critical temperature of $T_c=\lambda/2$. The surface
tension, $\sigma$, can be calculated analytically for a flat equilibrium
interface $\phi(y)$
orthogonal to the y direction as
$
\sigma=\kappa \int_{-\infty}^{\infty} (\partial_y \phi(y))^2 dy
$
where the equilibrium density profile of $\phi$ also depends on
$\kappa$.

\section{Viscoelasticity and the Boltzmann equation}
Viscoelasticity was first proposed by Maxwell in his dynamic theory
of gases\cite{maxwell}. He used the simple argument that in the limit
where there are no intermolecular collisions the fluid in a
container should behave like a solid: ``\dots Then it can easily be
shown that the pressures on the sides of the vessel due to the impacts
of the molecules are perfectly independent of each other, so that the
mass of moving molecules will behave, not like a fluid, but like a
solid.''  He goes on to deduce that the observed viscous behavior of
fluids is due to binary collisions that randomize the directions of
stress in the fluid. Since the collisions are fast, but not
instantaneous, the elastic properties of the fluid are not completely
lost, leading to the Maxwell model of viscoelasticity.

Subsequently, derivations of hydrodynamics from the dynamic theory of
gases have made the approximation of a purely viscous behavior
because of the difficulties of deriving a continuum approach at the
length scales of a mean free path of a molecule. In gases where lengths
less than the mean free path are important kinetic theory for rarefied 
gases is used.

There has recently been much activity in the research of the
experimentally observed viscoelastic behavior of simple liquids that
are undercooled. In this case, however, viscoelasticity is not
obtained because the relevant length scales were of the order of a mean
free path, but rather because of the correlations of subsequent
collisions as described in the mode coupling theory\cite{goetze}.

The arguments of Maxwell, however, are still valid for describing the
behavior of the Boltzmann equation, and viscoelastic properties can be
derived from the Boltzmann equation if the decay of viscous stresses
is slow.

The approach by Giraud {\it et al.} aims not at deriving a convected
Maxwell fluid, but a convected Jeffreys fluid which is a mixture
of a Maxwell fluid with a Newtonian fluid. A double set of
densities is introduced allowing two stresses, one of which is chosen
to relax quickly and is, therefore, a viscous stress, and the other,
which is chosen to decay very slowly, represents a viscoelastic
stress. The resulting model is a convected Jeffreys model that is often
used to describe a polymeric fluid in a solvent. Care has to be taken
for the choice of the collision matrix and the equilibrium
distribution to ensure an isotropic model. The details of this
one-component model are described in the publication by Giraud {\it et
al.}\cite{giraud}

A Chapman-Enskog expansion of the lattice Boltzmann equations
(\ref{feqn}) and (\ref{geqn}) gives the macroscopic equations that our
system simulates. Mass conservation gives the continuity equation:
\begin{equation}
\partial_t \rho + \partial_\alpha (\rho u_\alpha) = 0.
\end{equation}
Momentum conservation gives a Navier Stokes equation:
\begin{equation}
\rho \partial_t u_\alpha + \rho u_\beta \partial_\beta u_\alpha
= -\partial_\beta \left(P_{\alpha\beta}+\sigma^v_{\alpha\beta}+
\sigma_{\alpha\beta} \right) 
\end{equation}
where the viscous stress is given by
\begin{equation}
\sigma^v_{\alpha\beta}=\nu_\infty \partial_\beta (\rho u_\alpha)
+\xi_\infty \partial_\gamma (\rho u_\gamma) \delta_{\alpha\beta}.
\end{equation}
The  viscoelastic stress has the constitutive relation 
\begin{equation}\label{constitutive}
\sigma_{\alpha\beta} + \theta \sigma_{(1)\alpha\beta} = 
-(\nu_0-\nu_\infty) \left(\partial_\alpha (\rho u_\beta)+\partial_\beta
(\rho u_\alpha)\right)
\end{equation}
where $\sigma_{(1)}$ represents the upper convected derivative of
$\sigma$.  These equations are equivalent to the Navier-Stokes and
Jeffreys equations only in the incompressible limit where
$\partial_\alpha (\rho u_\beta)=\rho \partial_\alpha u_\beta$. The
fully compressible equations can only be simulated when a larger set
of velocities is used\cite{thesis}.
The conservation of the density difference leads to the convection
diffusion equation
\begin{equation}
\partial_t \phi + \partial_\alpha (\phi u_\alpha)
= D \partial_\alpha \partial_\alpha \mu+
\partial_\beta\left(\frac{\phi}{\rho} \partial_\alpha
(P_{\alpha\beta}-\sigma_{\alpha\beta})\right)
\end{equation}
where $D$ is a diffusion constant given by $D=(\tau-1/2)$.

\section{Simulation of a bubble in a viscoelastic liquid}
We applied our method to a system similar to the experimentally
well-studied system of an air bubble rising in a viscoelastic
fluid. In our simulation we represent the bubble using a
phase-separated Newtonian drop of low viscosity in matrix which is
viscoelastic by letting the relaxation time $\theta$ in equation
(\ref{constitutive}) depend smoothly on the density difference $\phi$
between $\theta=0.05$ in the drop and $\theta=66$ in the surrounding
fluid. We choose $\xi_\infty=0.06$, $\nu_\infty=0.01$, and
$\nu_0=0.01$ in the drop and $\nu_0=0.175$ outside. For the
thermodynamic parameters we select $T=0.5$, $\lambda=1.1$, and
$\kappa=0.007$, which corresponds to a surface tension of
$\sigma=0.02$. All units are in terms of the lattice spacings and the
time steps $\Delta t=\Delta x=1$. We introduce a forcing dependent on
$\phi$ so that the bubble is forced upward while the surrounding
fluid is forced downward. We choose the total change in the momentum
due to the forcing to be zero so that no walls are required in the
simulations.

We start the simulations without forcing and then periodically
increase the forcing after $10,000$ to $40,000$ iterations. We
observe the change in the velocity $u$ and store the distribution of
$\phi$ so that we have a way of judging the deformation of the bubble.


\begin{figure}
\begin{center}
\begin{minipage}{4cm}
\centerline{\psfig{figure=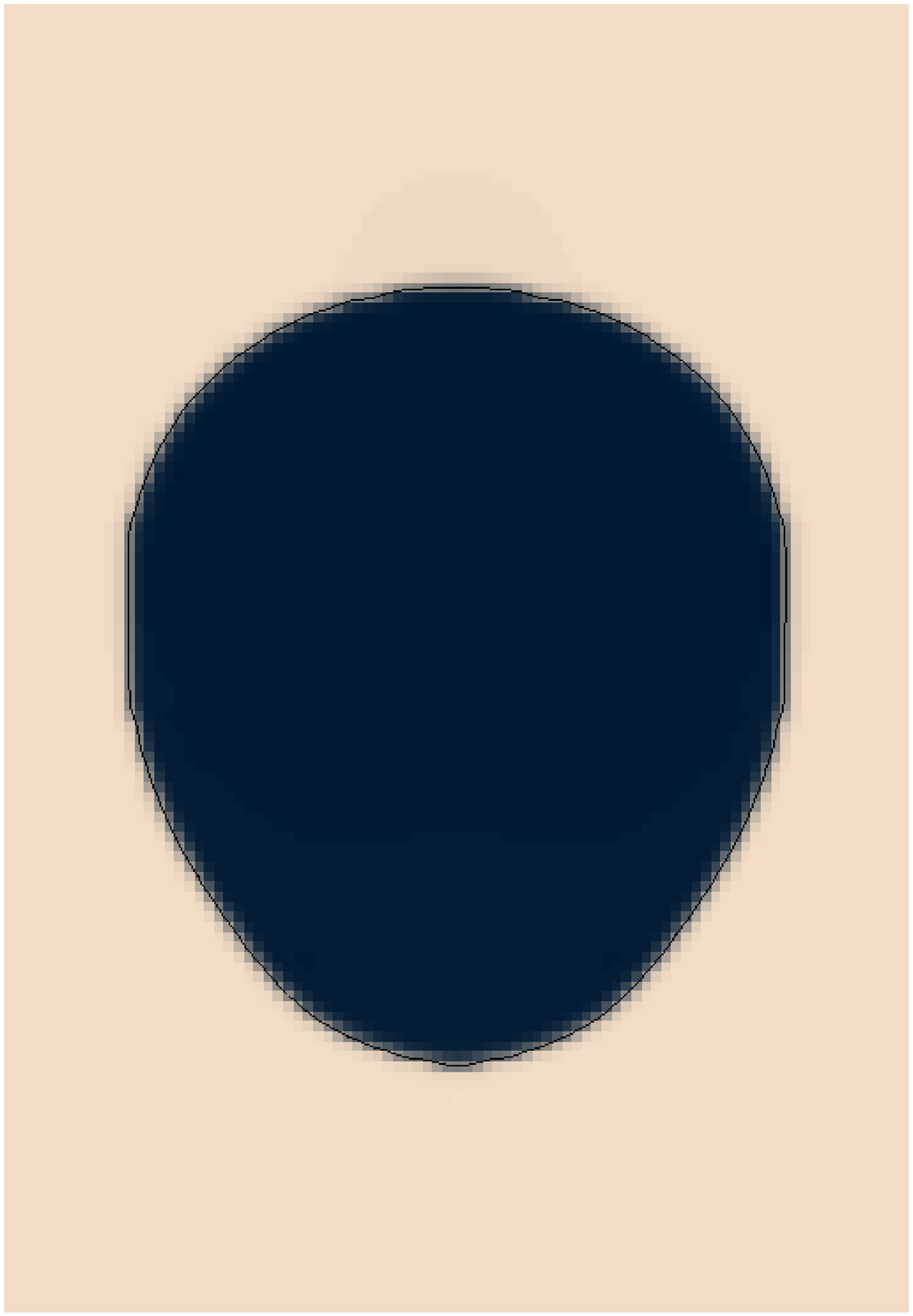,width=4cm}}
\begin{center} (a) \end{center}
\end{minipage}
\begin{minipage}{4cm}
\centerline{\psfig{figure=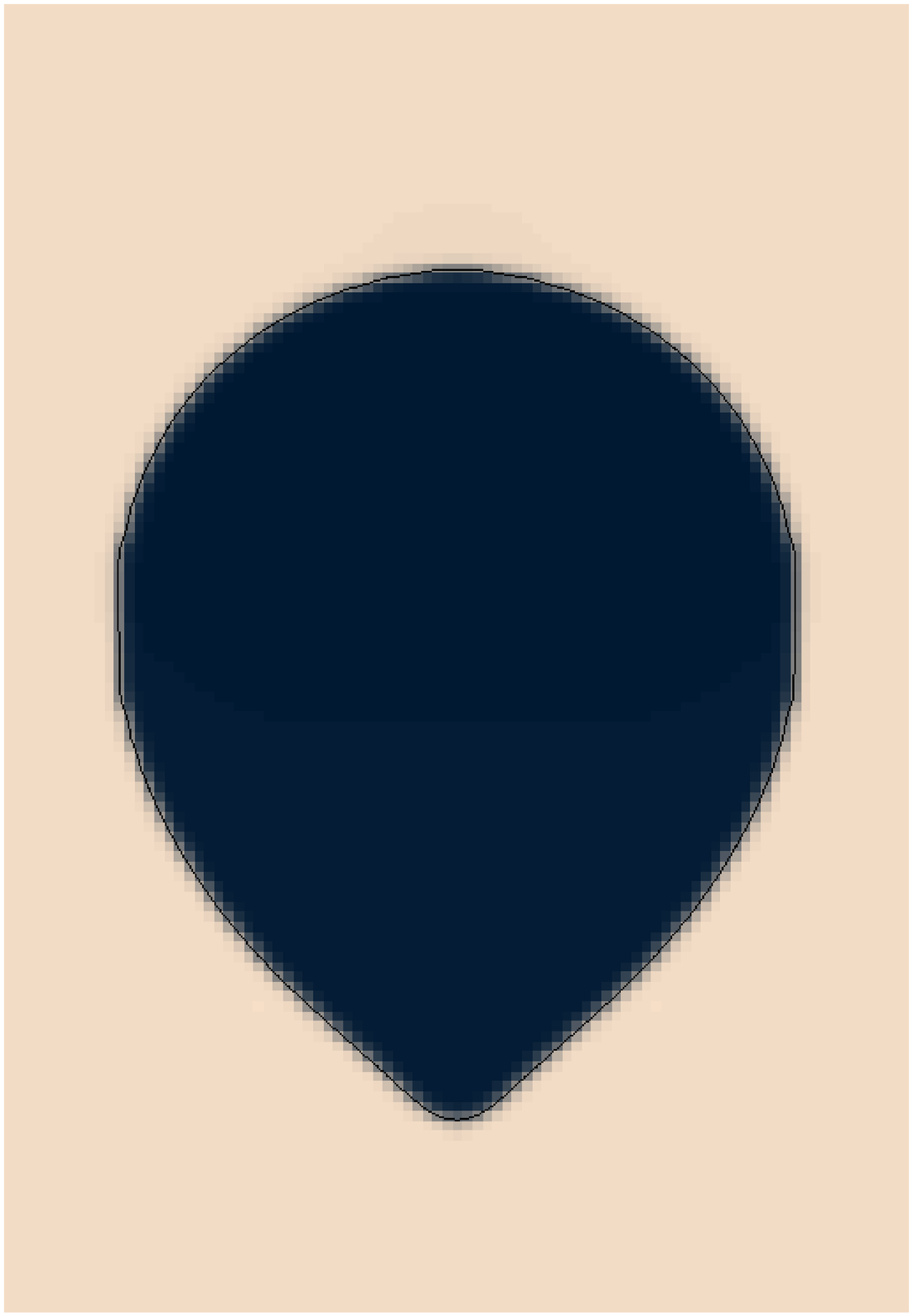,width=4cm}}
\begin{center} (b) \end{center}
\end{minipage}
\vspace{0.3cm}
\begin{minipage}{4cm}
\centerline{\psfig{figure=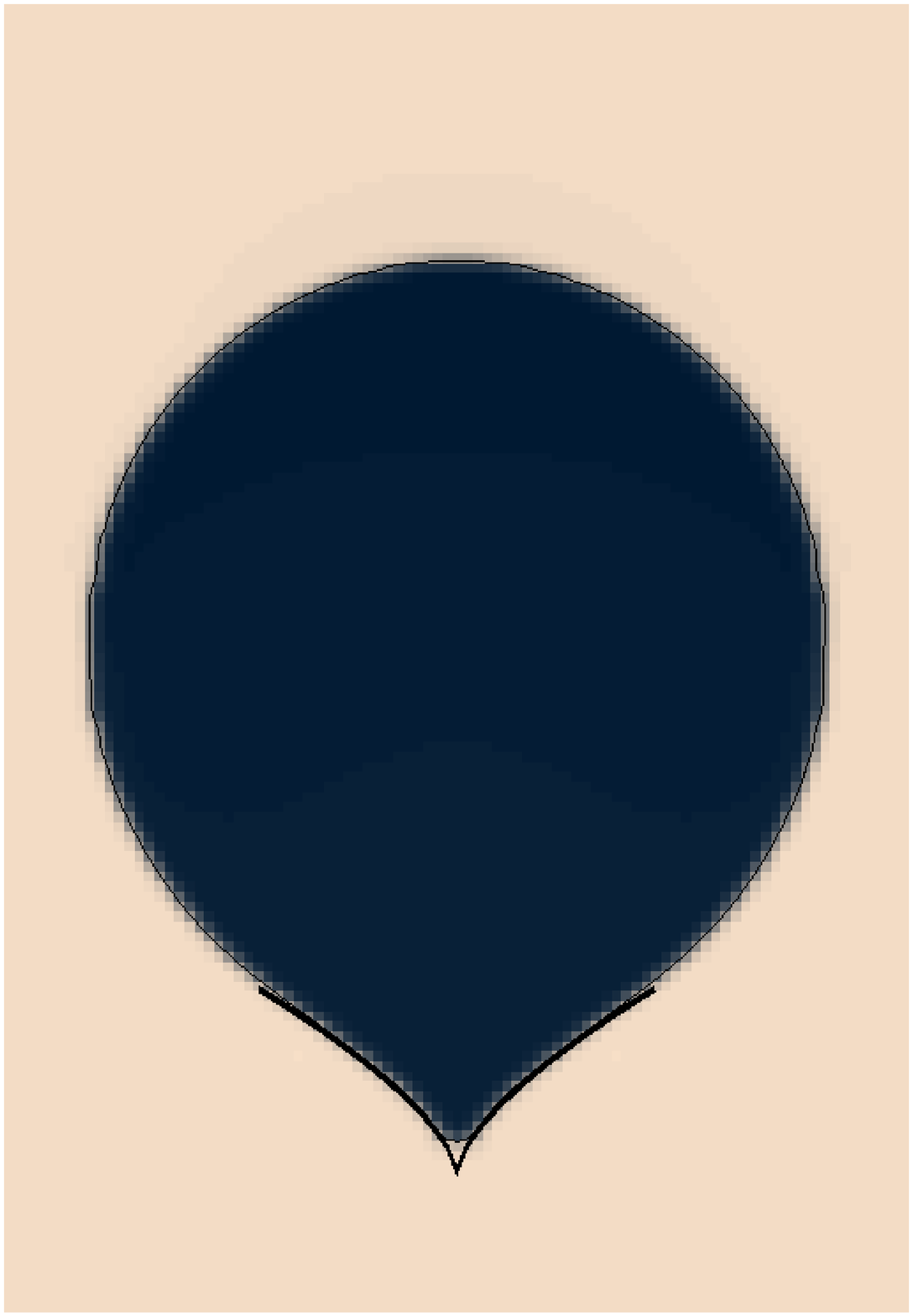,width=4cm}}
\begin{center} (c) \end{center}
\end{minipage}
\begin{minipage}{4cm}
\centerline{\psfig{figure=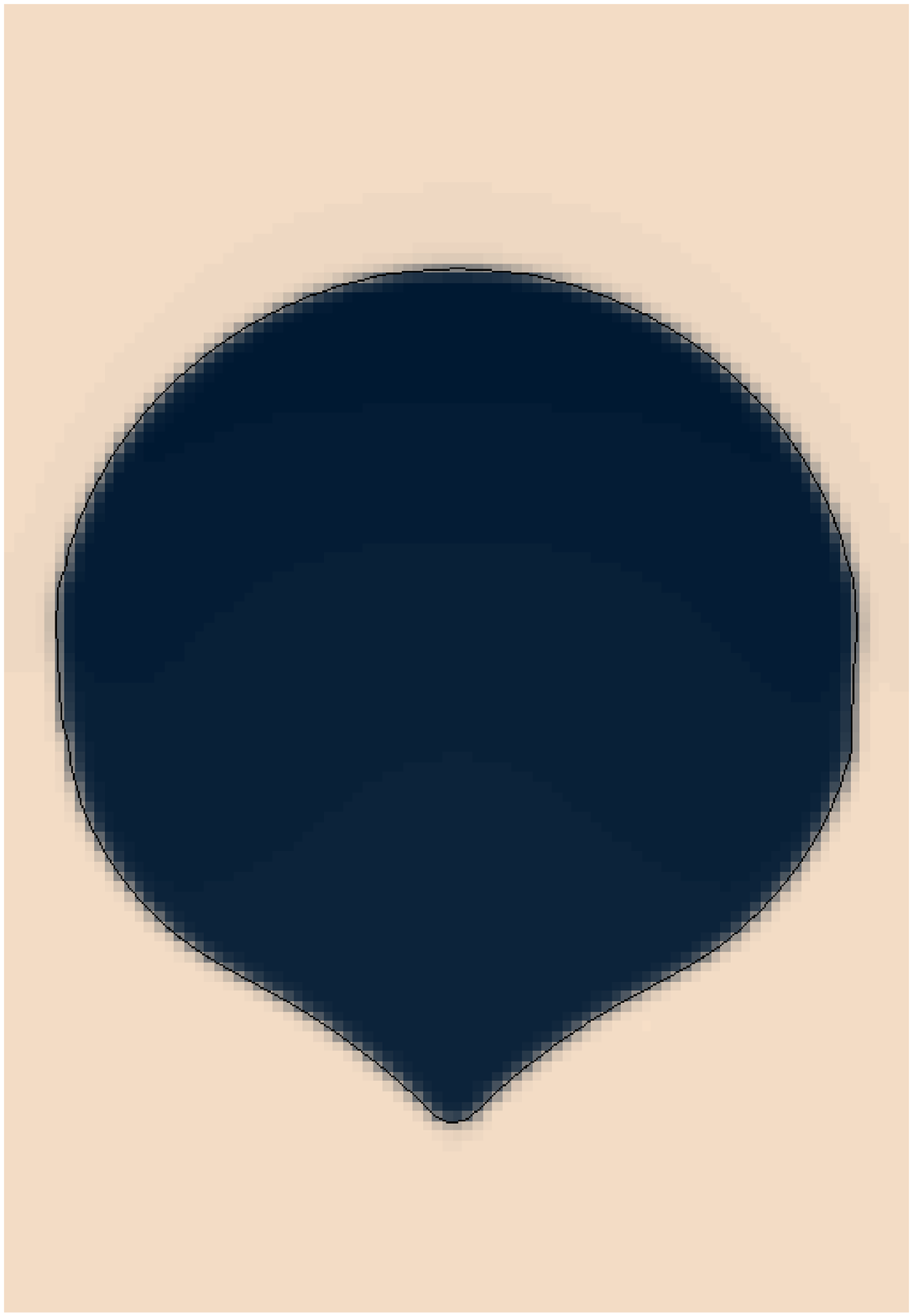,width=4cm}}
\begin{center} (d) \end{center}
\end{minipage}
\end{center}
\caption{ Shape for the simulated bubbles for different forcings in
(lattice spacings)/$\mbox{(time step)}^2$: (a)
$1.6 \cdot 10^{-5}$, (b) $3.6 \cdot 10^{-5}$, (c) $9.6 \cdot 10^{-5}$,
and (d) $1.96 \cdot 10^{-4}$. In (c) a fit to the predicted form of
the cusp ($|x|^{2/3}$ is also shown. All simulations are after a 20,000
iterations at this forcing.}
\label{fig:shapes}
\end{figure}

Figure \ref{fig:shapes} shows the form of the drop for different
forcings. At low forcings the drop is elongated in the flow
direction. This is in direct contrast to a bubble in a Newtonian
fluid, which is flattened in the flow direction. At a larger forcing
the bubble forms a cusp at the lower tip of the drop. For even larger
forcings the drop starts to flatten in the flow direction. This
sequence is in agreement with the experimental
findings\cite{liu}. The elongation of a rising bubble has been
simulated before\cite{noh}, but this is the first time that the
formation of a cusp has been simulated. In Figure \ref{fig:shapes}(c) it
is shown that the cusp can be fitted to the functional form $|x|{2/3}$
predicted by Joseph {\it et al.}\cite{joseph} for a two-dimensional
cusp created by the flow induced by two couter-rotating cylinders.

Experimentally the formation of a cusp has been observed to coincide
with a jump of nearly an order of magnitude in the terminal velocity
of the bubble\cite{bird,liu}, although the mechanism remains
disputed. On the one hand Bird {\it et al.}  argue that surface-active
impurities tend to immobilize bubble surfaces and hence retard the
motion of gas bubbles. This discontinuous change in bubble shape may
be responsible for the removal of the impurities, and thus lead to a
jump in the final velocity. Liu\cite{liu} {\it et al.} alternatively
suggest that the change in the shape of the bubble will make it more
streamlined, and therefore increase the terminal velocity.

\begin{figure}
\centerline{\psfig{figure=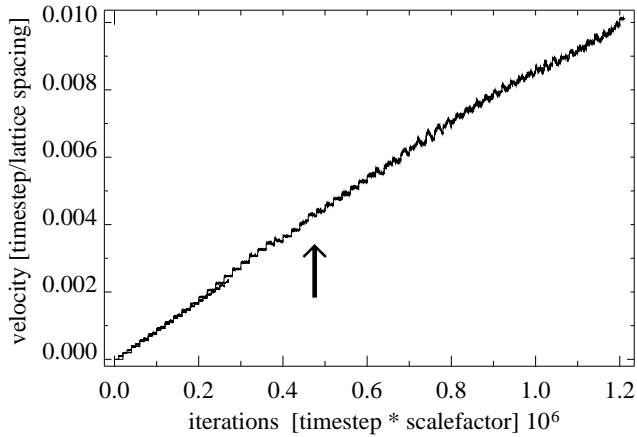,width=8.5cm}}
\caption{Velocity, u, of the bubble in two different sets of
simulations for a bubble of radius 35 in a 256x512 lattice. On the
x-axis the iterations were multiplied by a scale factor so that
corresponding forcings appear at the same point in the graph. No jump
in the final velocity is observed at $\sim 0.5\; 10^6$ iterations where the
formation of a cusp is observed (indicated by arrow). The scale
factors were 0.25 and 1. Velocity is measured in lattice units per
time step.}
\label{fig:uall}
\end{figure}

We examined the velocity for the rising drop as described above and
found no jump of about half an order of magnitude as observed by Liu
{\it et al.} in their experiment (see Figure \ref{fig:Bubble}).  Our
simulations suggest that the jump in velocity they observe is not
connected to a more streamline form of the bubble due to the cusp, but
more likely to the presence of surfactants that are absent in our
simulations.

\section{Conclusion}
We introduced a lattice Boltzmann model that can simulate viscoelastic
two-component flows.  We gave an intuitive explanation of the origin
of viscoelasticity in our model and the model by Giraud {\it et al.}
in terms of the original theory of Maxwell\cite{maxwell}.  Simulations
using this method have succeeded in reproducing the cusp at the end of
a bubble rising in a viscoelastic medium that have eluded earlier
numerical attempts with a more traditional boundary integral approach.

The model has been successful in the qualitative simulation of the
bubble problem in two dimensions. We intend to extend the model to
three dimensions in the future. This will also enable us to compare the
results quantitatively with experiment.

\section*{Acknowledgements}
One of us (A.W.) would like to thank Brad Chamberlain for his help in
implementing the algorithm in ZPL\cite{ZPL}. We would also like to
thank the Scientific Computing and Visualization center at Boston
University for a Mariner grant. L.G. is grateful for its support by
the ARC 97/02-210 project, Communaut\'e Fran{\c c}aise de Belgique.

\vspace{-0.7cm}

\def\jour#1#2#3#4{{#1} {\bf #2}, #3 (#4)}.
\def\tbp#1{{\em #1}, in preparation.}
\def\tit#1#2#3#4#5{{#1} {\bf #2}, #3 (#4).}
\def\ap{Adv. Phys.}
\def\epl{Euro. Phys. Lett.}
\def\pfa{Phys. Fluids A}
\def\prl{Phys. Rev. Lett.}
\def\pr{Phys. Rev.}
\def\pra{Phys. Rev. A}
\def\prb{Phys. Rev. B}
\def\pre{Phys. Rev. E}
\def\pa{Physica A}
\def\ps{Physica Scripta}
\def\zpb{Z. Phys. B}
\def\jmpc{J. Mod. Phys. C}
\def\jpc{J. Phys. C}
\def\jpcs{J. Phys. Chem. Solids}
\def\jpco{J. Phys. Cond. Mat}
\def\jf{J. Fluids}
\def\jfm{J. Fluid Mech.}
\def\arf{Ann. Rev. Fluid Mech.}
\def\roy{Proc. Roy. Soc.}
\def\rmp{Rev. Mod. Phys.}
\def\jsp{J. Stat. Phys.}
\def\pla{Phys. Lett. A}
\def\ijmp{Int. J. Mod. Phys. C}

\end{document}